\documentclass[pra,twocolumn,showpacs]{revtex4}
\usepackage{graphicx}

\begin{document}

\title{Controllable Frequency Entanglement via Auto-Phase-Matched Spontaneous Parametric Down-Conversion}

\author{Zachary~D.~Walton}
\email{walton@bu.edu} \homepage[Quantum Imaging Laboratory
homepage:~]{http://www.bu.edu/qil}

\author{Mark~C.~Booth}

\author{Alexander~V.~Sergienko}

\author{Bahaa~E.~A.~Saleh}

\author{Malvin~C.~Teich}
\affiliation{Quantum Imaging Laboratory, Department of Electrical
\& Computer Engineering, Boston University, 8 Saint Mary's Street,
Boston, Massachusetts 02215-2421}


\begin{abstract}

A new method for generating entangled photons with controllable
frequency correlation via spontaneous parametric down-conversion
(SPDC) is presented. The method entails initiating
counter-propagating SPDC in a single-mode nonlinear waveguide by
pumping with a pulsed beam perpendicular to the waveguide. The
method offers several advantages over other schemes, including the
ability to generate frequency-correlated photon pairs regardless
of the dispersion characteristics of the system.
\end{abstract}


\maketitle

Spontaneous parametric down-conversion (SPDC) is a convenient
process for generating pairs of photons that are entangled in one
or more of their respective degrees of freedom (direction,
frequency, polarization).  This entanglement can be used to
demonstrate counter-intuitive features of quantum mechanics and to
implement the growing suite of quantum information
technologies~\cite{Nielsen01}.  In a typical down-conversion
experiment, a photon from a monochromatic pump beam decays into
two photons (often referred to as signal and idler) via
interaction with a nonlinear optical crystal. While the signal and
idler may be broadband individually, conservation of energy
requires that the sum of their respective frequencies equals the
single frequency of the monochromatic pump. This engenders
frequency anti-correlation in the down-converted beams.  Aside
from the frequency-anti-correlated case, the frequency-correlated
and frequency-uncorrelated cases were also investigated
theoretically by Campos et al.~in 1990~\cite{Campos90}. At that
time, neither a method of creating these novel states nor a
practical application of the states was known.

Two developments in quantum information theory have renewed
interest in these generalized states of frequency correlation.
First, quantum information processes requiring the synchronized
creation of multiple photon pairs have been devised, such as
quantum teleportation~\cite{Bennett93} and entanglement
swapping~\cite{Pan98}.  The requisite temporal control can be
achieved by pumping the crystal with a brief pulse.  The
availability of pump photons of differing frequencies relaxes the
strict frequency anti-correlation in the down-converted
beams~\cite{Keller97}. Second, applications such as
entanglement-enhanced clock synchronization~\cite{Giovannetti01}
and one-way auto-compensating quantum
cryptography~\cite{Walton02a} have been introduced that
specifically require frequency correlation, as opposed to the
usual frequency anti-correlation.

\begin{figure}
\includegraphics{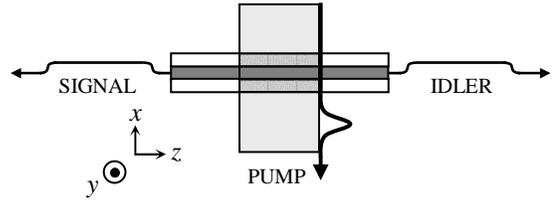}
\caption{A schematic of auto-phase-matched SPDC, a new method for
generating entangled-photon pairs with controllable frequency
correlation. The z-polarized pulsed pump beam initiates
counter-propagating y-polarized SPDC in the single-mode nonlinear
waveguide. The joint spectrum of the down-converted beams is
controlled by the spatial and temporal characteristics of the pump
beam, as described in the text.}\label{setup}
\end{figure}

Methods for preparing these novel states of frequency correlation
have emerged as well.  Keller and Rubin were first to observe that
when a specific relationship between the group velocities of the
pump and down-converted beams holds, the down-converted photons
are anti-correlated in time~\cite{Keller97}. Using a first-order
Taylor approximation of the relevant dispersion curves, they
provided two examples of bulk nonlinear crystals that satisfied
this relationship when used in a collinear type-II configuration
(signal and idler orthogonally polarized). Erdmann et al.~pointed
out that the time-anti-correlated state described by Keller and
Rubin entails frequency correlation (which can be seen immediately
by Fourier duality), and further emphasized that perfect frequency
correlation requires an infinite crystal, just as perfect
frequency anti-correlation requires a pump with an infinite
coherence length~\cite{Erdmann00}.  More recently, Giovannetti et
al.~demonstrated the feasibility of frequency-correlated
down-conversion in a periodically-poled nonlinear
crystal~\cite{Giovannetti02} and presented a formalism for
parameterizing the space of states between the cases of perfect
frequency correlation and perfect frequency
anti-correlation~\cite{Giovannetti02b}.


In this article, we present a new method for obtaining
controllable frequency entanglement that has distinct advantages
over the previously proposed methods.  Our method entails
initiating type-I SPDC (signal and idler identically polarized) in
a single-mode nonlinear waveguide by pumping with a pulsed beam
perpendicular to the waveguide (see Fig.~\ref{setup}). The
down-converted photons emerge from opposite ends of the waveguide
with a joint spectrum that can be varied from frequency
anti-correlated to frequency correlated by adjusting the temporal
and spatial characteristics of the pump beam. The primary
advantage of this method is that the limiting cases of perfect
frequency correlation and perfect frequency anti-correlation can
be obtained regardless of the dispersion relation of the
waveguide. Thus, we refer to the method as
\textit{auto-phase-matched}. It is well known that the
frequency-anti-correlated case is achievable regardless of the
dispersion relations in a collinear configuration with a
monochromatic pump and a thin bulk crystal; however, the
frequency-correlated case has hitherto been associated with a
constraint on the dispersion relations (cf.~the ``group velocity
matching'' condition introduced in Ref.~\cite{Keller97}). The
geometry we propose restores the symmetry between the two cases by
ensuring the appropriate phase-matching regardless of the
dispersion relation of the waveguide.

This article is organized as follows.  First, we write the output
state of the SPDC produced in our new configuration and provide an
estimate of the conversion efficiency. Second, we analyze the
state using a Franson interferometer~\cite{Franson89}, which
illustrates the duality between frequency correlation and
anti-correlation. Third, we quantify our method's advantage by
comparing the visibility achieved in a Franson interferometer by
the frequency-correlated, collinear configuration described in
Ref.~\cite{Giovannetti02} with the visibility achieved in our
counter-propagating configuration.

The transverse-pump, counter-propagating geometry depicted in
Fig.~\ref{setup} has been noted as a promising source of
entangled-photon pairs for both type-I~\cite{Booth02} and
type-II~\cite{DeRossi02} SPDC.  The most obvious advantages of
this geometry over a collinear geometry pertain to the separation
of the three interacting beams. In a transverse-pump,
counter-propagating geometry, all three beams are traveling in
different directions.  Thus, the usual techniques for filtering
the pump beam from the down-conversion and separating the
down-converted beams at a beamsplitter are unnecessary.

The investigations in Refs.~\cite{Booth02,DeRossi02} were limited
to the case of a monochromatic pump beam.  There are two primary
advantages of pumping with a broadband beam perpendicular to the
waveguide and arranging for type-I down-conversion. First, the
dispersion relation for the pump beam plays no role in the
phase-matching analysis, since the waveguide ensures
phase-matching in the transverse direction. Second, the
counter-propagating, identically-polarized signal and idler fields
will be phase-matched in the long-crystal limit only if they have
equal and opposite propagation vectors, a condition which entails
equal frequency.  Thus, the bandwidth of the pump determines the
allowable range of the sum frequency of the signal and idler, and
the longitudinal length of the illuminated portion of the crystal
determines the allowable range of the difference frequency.

We assume that the nonlinear coefficient and the propagation
constants vary sufficiently slowly with frequency that they may be
taken outside any frequency integrals in which they appear as
integrand prefactors.  Furthermore, we assume that the waveguide
is long compared to the width of the pump beam such that the
interaction length is controlled by the pump beam profile along
the z-axis (see Fig.~\ref{setup}).  Following the derivation in
Ref.~\cite{Booth02} of the quantum state of a counter-propagating
photon pair, we have
\begin{equation}\label{Psi}
|\Psi\rangle \propto \int\!\!\!\int d\omega_{l}\, d\omega_r\,
\tilde{E_t}(\omega_l+\omega_r)\tilde{f_z}\left[\Delta\beta(\omega_l,\omega_r)\right]\,|\omega_l\rangle_l\,|\omega_r\rangle_r,
\end{equation}
where $\tilde{E_t}(\omega)$ and $\tilde{f_z}(\Delta\beta)$ are the
respective Fourier transforms of the temporal and spatial
functions describing the pump beam $E_p(t,z)=E_t(t)f_z(z)$,
$\Delta\beta(\omega_l,\omega_r)=\beta(\omega_l)-\beta(\omega_r)$
is the difference in the waveguide propagation constant evaluated
at $\omega_l$ and $\omega_r$, and $|\omega\rangle_{l(r)}$ denotes
a single photon at frequency $\omega$ moving to the left(right).


To investigate the dependence of $|\Psi\rangle$ on the
characteristics of the pump, we choose Gaussian profiles in space
and time for the pump pulse, such that $\tilde{E_t}(\omega)\propto
e^{-\frac{1}{2}(\omega\tau)^2}$ and
$\tilde{f_z}(\Delta\beta)\propto e^{-\frac{1}{2}(\Delta\beta
W)^2}$, where $\tau$ and $W$ are the duration and width (along the
$z$-axis in Fig.~\ref{setup}) of the pump pulse, respectively. In
the limit of a monochromatic pump ($\tau\rightarrow\infty$) with
finite spatial extent, $\tilde{E_t}(\omega_l+\omega_r)$ is sharply
peaked around the pump center frequency. Thus, the sum frequency
of the signal and idler is fixed. This is the familiar
frequency-anti-correlated case that is readily achievable in thin
bulk crystals. In the limit of a finite-duration pump pulse of
infinite spatial extent ($W\rightarrow\infty$),
$\tilde{f_z}\left[\Delta\beta(\omega_l,\omega_r)\right]$ is
sharply peaked around $\Delta\beta=0$.  Thus, photon pairs for
which $\omega_l\approx\omega_r$ predominate.  This is the
frequency-correlated case that has hitherto only been achieved by
imposing a group velocity matching condition.

The efficiency of this geometry in an GaAs-based waveguide of
length 1 mm and transverse dimension 3 $\mu\textrm{m}$ is
calculated in Ref.~\cite{DeRossi02} to range between $10^{-9}$ and
$10^{-11}$ depending on the transverse profile of the waveguide.
These figures compare favorably with the SPDC efficiencies
achieved in more conventional bulk-crystal configurations (e.g.,
$10^{-13}$ in Ref.~\cite{Jennewein00}), though they are still
several orders of magnitude less than that achieved in
periodically poled lithium niobate waveguides (e.g., $10^{-6}$ in
Ref.~\cite{Tanzilli01_EL}).

\begin{figure}
\includegraphics{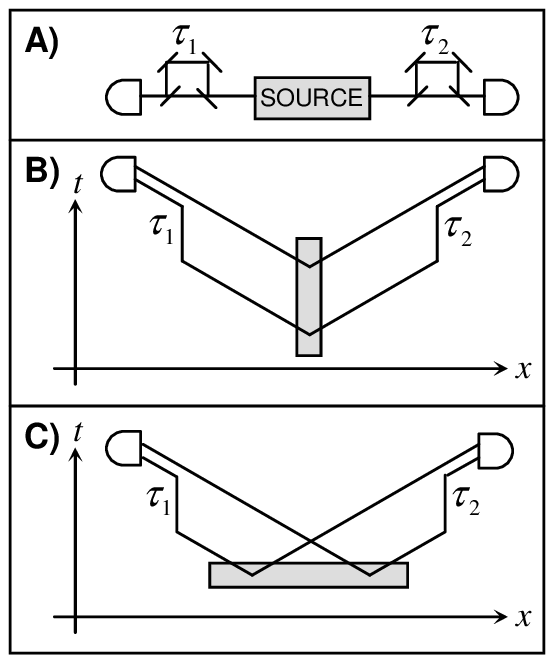}
\caption{The Franson interferometer (A) and the two types of
indistinguishability it can bring about (B and C). B) depicts the
indistinguishability in time of creation of the photon pair, and
C) depicts indistinguishability in position of creation of the
photon pair. These two cases are shown in the text to correspond
to frequency-anti-correlated photon pairs and frequency-correlated
photon pairs, respectively.}\label{duality}
\end{figure}

The Franson interferometer~\cite{Franson89} is a natural tool for
distinguishing frequency correlation and frequency
anti-correlation. When the two delays ($\tau_1$ and $\tau_2$) are
equal to within the reciprocal bandwidth of down-conversion,
coincidence detections can be associated with indistinguishable
pair-creation events (see Fig.~\ref{duality}A). If the
down-converted photons are correlated in time (anti-correlated in
frequency), the short-short two-photon amplitude interferes with
the long-long amplitude (see Fig.~\ref{duality}B).  If the
down-converted photons are anti-correlated in time (correlated in
frequency), the short-long amplitude interferes with the
long-short amplitude (see Fig.~\ref{duality}C).  The duality
between these two cases can be seen by comparing the loci of
indistinguishable pair-creation events in the spacetime diagrams
of Fig.~\ref{duality}B and Fig.~\ref{duality}C. The
frequency-anti-correlated case depicted in Fig.~\ref{duality}B
arises from the coherent superposition of pair-creation events at
a fixed position over a range of times, while the
frequency-correlated case depicted in Fig.~\ref{duality}C arises
from the coherent superposition of pair-creation events at a fixed
time over a range of positions. Note that while the interference
visibility decreases in both cases as $\tau_1-\tau_2$ approaches
the reciprocal bandwidth of down-conversion, the relative phase
between the interfering amplitudes depends on $\tau_1+\tau_2$ in
the frequency-anti-correlated case, and on $\tau_1-\tau_2$ in the
frequency-correlated case.


\begin{figure}
\includegraphics{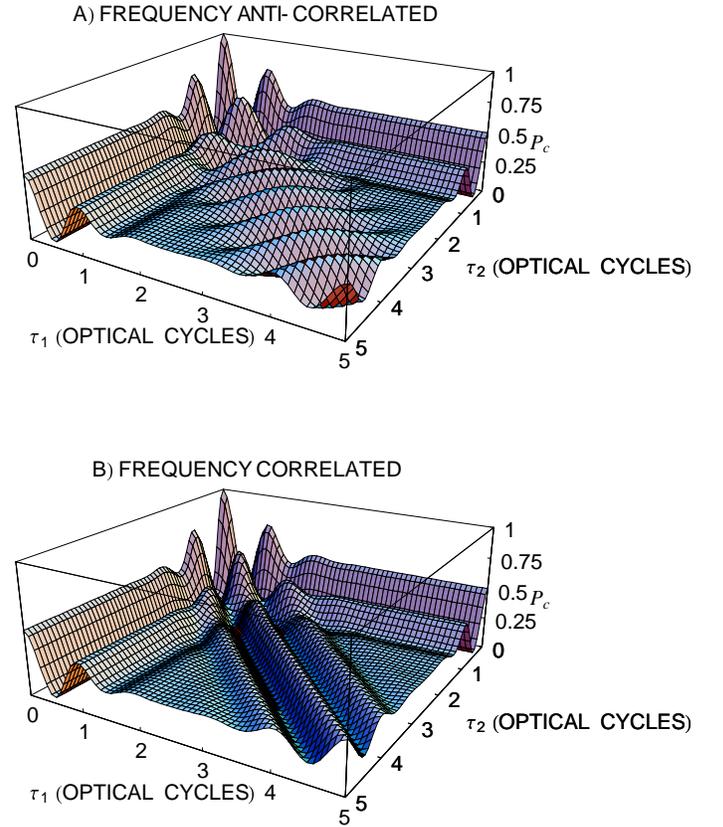}
\caption{The probability of coincidence when
frequency-anti-correlated (A) and frequency-correlated (B) states
are analyzed with a Franson interferometer (see
Fig.~\protect\ref{duality}). The down-converted beams have center
frequency $\omega_p/2$ and bandwidth $\omega_p/10$. $\tau_1$ and
$\tau_2$ are in units of optical cycles at the center frequency of
downconversion.}\label{math}
\end{figure}

In Fig.~\ref{math} we plot the probability of coincidence in the
Franson interferometer for the aforementioned limiting cases of
the two-photon source: perfect frequency anti-correlation
($\tau\rightarrow\infty$, $W\rightarrow\,$finite), and perfect
frequency-anti-correlation ($\tau\rightarrow\,$finite,
$W\rightarrow\infty$). The finite values of $\tau$ and $W$ are
chosen such that the bandwidth of downconversion is $\omega_p/10$
in each case. The fourth-order fringes in the
$\tau_1\approx\tau_2\gg10/\omega_p$ region show that the Franson
interferometer clearly distinguishes the two cases. The modulation
is in the $\Delta\tau_1=\Delta\tau_2$ direction the
frequency-anti-correlated case and in the
$\Delta\tau_1=-\Delta\tau_2$ direction in the frequency-correlated
case.

\begin{figure}
\includegraphics{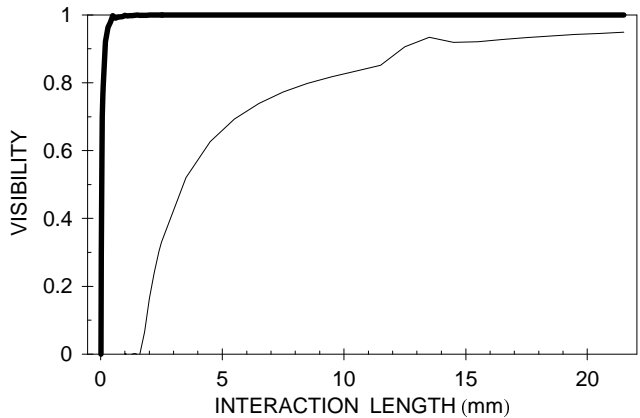}
\caption{Numerical calculation of the fourth-order fringe
visibility seen in a Franson interferometer when the perfect
source of frequency-correlated photon pairs is approximated by the
method described in Ref.~\protect\cite{Giovannetti02} (thin line)
and by the auto-phase-matched method described in the text (bold
line).  The plot depicts the effect of changing the interaction
length of the nonlinear process while holding the bandwidth of the
pump fixed.}\label{visibility}
\end{figure}

By establishing the signature of the perfect frequency-correlated
state (the fourth-order fringes in Fig.~\ref{math}B), we are able
to compare the performance of experimental methods designed to
produce this state.  Specifically, the visibility of the fringes
in Fig.~\ref{math}B provides a measure of the quality of the
frequency-correlated state.  In Fig.~\ref{visibility} we plot a
numerical calculation of the visibility achieved by the source
described in Ref.~\cite{Giovannetti02} (thin line) and that
achieved by our auto-phase-matched method (thick line) in a
GaAs-based waveguide, for a range of interaction lengths.  In the
method of Ref.~\cite{Giovannetti02} the interaction length is the
thickness of the crystal, while in our method the interaction
length is the width of the pump beam in along the z-axis (see
Fig.~\ref{setup}). In order to minimize the complicating effect of
second-order interference, the visibility is calculated at the
delay offset $(\tau_1,\tau_2)=(4/\sigma,4/\sigma)$ where $\sigma$
is the bandwidth of down-conversion.

While the fourth-order visibility in Fig.~\ref{math}B is 0.5, we
have scaled the visibilities to range between 0 and 1 in
Fig.~\ref{visibility} since methods exist for restoring maximal
visibility in the Franson interferometer.  The first method put
forward involves time-gating the detectors~\cite{Brendel91}. For
that method to work with the source described here, the waveguide
would have to be much longer ($>$5 cm) than those in current
experimental designs ($\sim$3 mm in Ref.~\cite{DeRossi01}). A more
promising approach for restoring visibility exploits polarization
entanglement, as demonstrated in Ref.~\cite{Strekalov96}. At the
end of this paper we discuss the natural way in which the source
described here enables concurrent polarization and frequency
entanglement.

The parameters of the source described in
Ref.~\cite{Giovannetti02} are as follows: periodically-poled
potassuium titanyl phosphate (PPKTP) with a poling period of 47.7
$\mu\textrm{m}$, pump wavelength of 790 nm, pump bandwidth of 3
THz,  and collinear propagation along the crystal's $x$ axis with
the signal $z$-polarized and the pump and idler $y$-polarized. For
our new geometry, we consider a GaAs slab waveguide which is 3
$\mu\textrm{m}$ in the transverse dimension and configured as
depicted in Fig.~\ref{setup}, with the pump $z$-polarized and the
signal and idler polarized in the $x$-$y$ plane.  Since both
schemes rely on a long-crystal phase-matching condition, it is not
surprising that the visibility of each increases with increasing
interaction length; however, while the deleterious influence of
higher-order terms in the crystal dispersion relations is
exacerbated as the crystal length is increased in the method of
Ref.~\cite{Giovannetti02}, our auto-phase-matched approach
provides an increasingly close approximation to the desired state,
regardless of the dispersion relations.


Finally, it is worth noting that our source may be used to create
photon pairs that are entangled in polarization. In
Fig.~\ref{setup}, we were only concerned with down-converted
photons polarized along the y-axis. If the pump beam is polarized
along the z-axis and we use a material with the symmetry
properties of GaAs which has $\chi^{(2)}_{zxy}=\chi^{(2)}_{zyx}$
and $\chi^{(2)}_{zxx}=\chi^{(2)}_{zyy}=0$~\cite{Yariv88}, we will
get a counter-propagating polarization-entangled state
($|HV\rangle+|VH\rangle$ directly from the crystal~\footnote{The
relative phase between the two terms ($|HV\rangle$ and
$|VH\rangle$) in the superposition is unity since GaAs is not
birefringent.}. Furthermore, we can obtain this
polarization-entanglement while independently controlling the
frequency entanglement by manipulating the pump beam, as
previously described.

\begin{acknowledgments}
This work was supported by the National Science Foundation; the
Center for Subsurface Sensing and Imaging Systems (CenSSIS), an
NSF Engineering Research Center; the Defense Advanced Research
Projects Agency (DARPA); and the David and Lucile Packard
Foundation.

\end{acknowledgments}


\begin{thebibliography}{19}
\expandafter\ifx\csname
natexlab\endcsname\relax\def\natexlab#1{#1}\fi
\expandafter\ifx\csname bibnamefont\endcsname\relax
  \def\bibnamefont#1{#1}\fi
\expandafter\ifx\csname bibfnamefont\endcsname\relax
  \def\bibfnamefont#1{#1}\fi
\expandafter\ifx\csname citenamefont\endcsname\relax
  \def\citenamefont#1{#1}\fi
\expandafter\ifx\csname url\endcsname\relax
  \def\url#1{\texttt{#1}}\fi
\expandafter\ifx\csname
urlprefix\endcsname\relax\def\urlprefix{URL }\fi
\providecommand{\bibinfo}[2]{#2}
\providecommand{\eprint}[2][]{\url{#2}}

\bibitem[{\citenamefont{Nielsen and Chuang}(2001)}]{Nielsen01}
\bibinfo{author}{\bibfnamefont{M.~A.} \bibnamefont{Nielsen}} \bibnamefont{and}
  \bibinfo{author}{\bibfnamefont{I.~L.} \bibnamefont{Chuang}},
  \emph{\bibinfo{title}{Quantum Computing and Quantum Information}}
  (\bibinfo{publisher}{Cambridge}, \bibinfo{address}{New York},
  \bibinfo{year}{2001}).

\bibitem[{\citenamefont{Campos et~al.}(1990)\citenamefont{Campos, Saleh, and
  Teich}}]{Campos90}
\bibinfo{author}{\bibfnamefont{R.~A.} \bibnamefont{Campos}},
  \bibinfo{author}{\bibfnamefont{B.~E.~A.} \bibnamefont{Saleh}},
  \bibnamefont{and} \bibinfo{author}{\bibfnamefont{M.~C.} \bibnamefont{Teich}},
  \bibinfo{journal}{Phys. Rev. A} \textbf{\bibinfo{volume}{42}},
  \bibinfo{pages}{4127} (\bibinfo{year}{1990}).

\bibitem[{\citenamefont{Bennett et~al.}(1993)\citenamefont{Bennett, Brassard,
  Cr\'{e}peau, Jozsa, Peres, and Wootters}}]{Bennett93}
\bibinfo{author}{\bibfnamefont{C.~H.} \bibnamefont{Bennett}},
  \bibinfo{author}{\bibfnamefont{G.}~\bibnamefont{Brassard}},
  \bibinfo{author}{\bibfnamefont{C.}~\bibnamefont{Cr\'{e}peau}},
  \bibinfo{author}{\bibfnamefont{R.}~\bibnamefont{Jozsa}},
  \bibinfo{author}{\bibfnamefont{A.}~\bibnamefont{Peres}}, \bibnamefont{and}
  \bibinfo{author}{\bibfnamefont{W.~K.} \bibnamefont{Wootters}},
  \bibinfo{journal}{Phys. Rev. Lett.} \textbf{\bibinfo{volume}{70}},
  \bibinfo{pages}{1895} (\bibinfo{year}{1993}).

\bibitem[{\citenamefont{Pan et~al.}(1998)\citenamefont{Pan, Bouwmeester,
  Weinfurter, and Zeilinger}}]{Pan98}
\bibinfo{author}{\bibfnamefont{J.~W.} \bibnamefont{Pan}},
  \bibinfo{author}{\bibfnamefont{D.}~\bibnamefont{Bouwmeester}},
  \bibinfo{author}{\bibfnamefont{H.}~\bibnamefont{Weinfurter}},
  \bibnamefont{and}
  \bibinfo{author}{\bibfnamefont{A.}~\bibnamefont{Zeilinger}},
  \bibinfo{journal}{Phys. Rev. Lett.} \textbf{\bibinfo{volume}{80}},
  \bibinfo{pages}{3891} (\bibinfo{year}{1998}).

\bibitem[{\citenamefont{Keller and Rubin}(1997)}]{Keller97}
\bibinfo{author}{\bibfnamefont{T.~E.} \bibnamefont{Keller}} \bibnamefont{and}
  \bibinfo{author}{\bibfnamefont{M.~H.} \bibnamefont{Rubin}},
  \bibinfo{journal}{Phys. Rev. A} \textbf{\bibinfo{volume}{56}},
  \bibinfo{pages}{1534} (\bibinfo{year}{1997}).

\bibitem[{\citenamefont{Giovannetti
  et~al.}(2002{\natexlab{a}})\citenamefont{Giovannetti, Lloyd, and
  Maccone}}]{Giovannetti01}
\bibinfo{author}{\bibfnamefont{V.}~\bibnamefont{Giovannetti}},
  \bibinfo{author}{\bibfnamefont{S.}~\bibnamefont{Lloyd}}, \bibnamefont{and}
  \bibinfo{author}{\bibfnamefont{L.}~\bibnamefont{Maccone}},
  \bibinfo{journal}{Nature} \textbf{\bibinfo{volume}{412}},
  \bibinfo{pages}{417} (\bibinfo{year}{2002}{\natexlab{a}}).

\bibitem[{\citenamefont{Walton et~al.}(2002)\citenamefont{Walton, Abouraddy,
  Sergienko, Saleh, and Teich}}]{Walton02a}
\bibinfo{author}{\bibfnamefont{Z.}~\bibnamefont{Walton}},
  \bibinfo{author}{\bibfnamefont{A.~F.} \bibnamefont{Abouraddy}},
  \bibinfo{author}{\bibfnamefont{A.~V.} \bibnamefont{Sergienko}},
  \bibinfo{author}{\bibfnamefont{B.~E.~A.} \bibnamefont{Saleh}},
  \bibnamefont{and} \bibinfo{author}{\bibfnamefont{M.~C.} \bibnamefont{Teich}},
  \bibinfo{journal}{quant-ph/0207167}  (\bibinfo{year}{2002}).

\bibitem[{\citenamefont{Erdmann et~al.}(2000)\citenamefont{Erdmann, Branning,
  Grice, and Walmsley}}]{Erdmann00}
\bibinfo{author}{\bibfnamefont{R.}~\bibnamefont{Erdmann}},
  \bibinfo{author}{\bibfnamefont{D.}~\bibnamefont{Branning}},
  \bibinfo{author}{\bibfnamefont{W.}~\bibnamefont{Grice}}, \bibnamefont{and}
  \bibinfo{author}{\bibfnamefont{I.~A.} \bibnamefont{Walmsley}},
  \bibinfo{journal}{Phys. Rev. A} \textbf{\bibinfo{volume}{62}},
  \bibinfo{pages}{53810} (\bibinfo{year}{2000}).

\bibitem[{\citenamefont{Giovannetti
  et~al.}(2002{\natexlab{b}})\citenamefont{Giovannetti, Maccone, Shapiro, and
  Wong}}]{Giovannetti02}
\bibinfo{author}{\bibfnamefont{V.}~\bibnamefont{Giovannetti}},
  \bibinfo{author}{\bibfnamefont{L.}~\bibnamefont{Maccone}},
  \bibinfo{author}{\bibfnamefont{J.~H.} \bibnamefont{Shapiro}},
  \bibnamefont{and} \bibinfo{author}{\bibfnamefont{F.~N.~C.}
  \bibnamefont{Wong}}, \bibinfo{journal}{Phys. Rev. Lett.}
  \textbf{\bibinfo{volume}{88}}, \bibinfo{pages}{183602}
  (\bibinfo{year}{2002}{\natexlab{b}}).

\bibitem[{\citenamefont{Giovannetti
  et~al.}(2002{\natexlab{c}})\citenamefont{Giovannetti, Maccone, Shapiro, and
  Wong}}]{Giovannetti02b}
\bibinfo{author}{\bibfnamefont{V.}~\bibnamefont{Giovannetti}},
  \bibinfo{author}{\bibfnamefont{L.}~\bibnamefont{Maccone}},
  \bibinfo{author}{\bibfnamefont{J.~H.} \bibnamefont{Shapiro}},
  \bibnamefont{and} \bibinfo{author}{\bibfnamefont{F.~N.~C.}
  \bibnamefont{Wong}}, \bibinfo{journal}{Phys. Rev. A}
  \textbf{\bibinfo{volume}{66}}, \bibinfo{pages}{43813}
  (\bibinfo{year}{2002}{\natexlab{c}}).

\bibitem[{\citenamefont{Franson}(1989)}]{Franson89}
\bibinfo{author}{\bibfnamefont{J.~D.} \bibnamefont{Franson}},
  \bibinfo{journal}{Phys. Rev. Lett.} \textbf{\bibinfo{volume}{62}},
  \bibinfo{pages}{2205} (\bibinfo{year}{1989}).

\bibitem[{\citenamefont{Booth et~al.}(2002)\citenamefont{Booth, Atat{\" u}re,
  {Di Giuseppe}, Saleh, Sergienko, and Teich}}]{Booth02}
\bibinfo{author}{\bibfnamefont{M.~C.} \bibnamefont{Booth}},
  \bibinfo{author}{\bibfnamefont{M.}~\bibnamefont{Atat{\" u}re}},
  \bibinfo{author}{\bibfnamefont{G.}~\bibnamefont{{Di Giuseppe}}},
  \bibinfo{author}{\bibfnamefont{B.~E.~A.} \bibnamefont{Saleh}},
  \bibinfo{author}{\bibfnamefont{A.}~\bibnamefont{Sergienko}},
  \bibnamefont{and} \bibinfo{author}{\bibfnamefont{M.~C.} \bibnamefont{Teich}},
  \bibinfo{journal}{Phys. Rev. A} \textbf{\bibinfo{volume}{66}},
  \bibinfo{pages}{23815} (\bibinfo{year}{2002}).

\bibitem[{\citenamefont{{De Rossi} and Berger}(2002)}]{DeRossi02}
\bibinfo{author}{\bibfnamefont{A.}~\bibnamefont{{De Rossi}}} \bibnamefont{and}
  \bibinfo{author}{\bibfnamefont{V.}~\bibnamefont{Berger}},
  \bibinfo{journal}{Phys. Rev. Lett.} \textbf{\bibinfo{volume}{88}},
  \bibinfo{pages}{043901} (\bibinfo{year}{2002}).

\bibitem[{\citenamefont{Jennewein et~al.}(2000)\citenamefont{Jennewein, Simon,
  Weihs, Weinfurter, and Zeilinger}}]{Jennewein00}
\bibinfo{author}{\bibfnamefont{T.}~\bibnamefont{Jennewein}},
  \bibinfo{author}{\bibfnamefont{C.}~\bibnamefont{Simon}},
  \bibinfo{author}{\bibfnamefont{G.}~\bibnamefont{Weihs}},
  \bibinfo{author}{\bibfnamefont{H.}~\bibnamefont{Weinfurter}},
  \bibnamefont{and}
  \bibinfo{author}{\bibfnamefont{A.}~\bibnamefont{Zeilinger}},
  \bibinfo{journal}{Phys. Rev. Lett.} \textbf{\bibinfo{volume}{84}},
  \bibinfo{pages}{4729} (\bibinfo{year}{2000}).

\bibitem[{\citenamefont{Tanzilli et~al.}(2001)\citenamefont{Tanzilli,
  Riedmatten, Tittel, Zbinden, Baldi, Micheli, Ostrowsky, and
  Gisin}}]{Tanzilli01_EL}
\bibinfo{author}{\bibfnamefont{S.}~\bibnamefont{Tanzilli}},
  \bibinfo{author}{\bibfnamefont{H.~D.} \bibnamefont{Riedmatten}},
  \bibinfo{author}{\bibfnamefont{W.}~\bibnamefont{Tittel}},
  \bibinfo{author}{\bibfnamefont{H.}~\bibnamefont{Zbinden}},
  \bibinfo{author}{\bibfnamefont{P.}~\bibnamefont{Baldi}},
  \bibinfo{author}{\bibfnamefont{M.~D.} \bibnamefont{Micheli}},
  \bibinfo{author}{\bibfnamefont{D.~B.} \bibnamefont{Ostrowsky}},
  \bibnamefont{and} \bibinfo{author}{\bibfnamefont{N.}~\bibnamefont{Gisin}},
  \bibinfo{journal}{Electron. Lett.} \textbf{\bibinfo{volume}{37}},
  \bibinfo{pages}{26} (\bibinfo{year}{2001}).

\bibitem[{\citenamefont{Yariv}(1988)}]{Yariv88}
\bibinfo{author}{\bibfnamefont{A.}~\bibnamefont{Yariv}},
  \emph{\bibinfo{title}{Quantum Electronics}} (\bibinfo{publisher}{John Wiley
  \& Sons}, \bibinfo{address}{New York}, \bibinfo{year}{1988}),
  \bibinfo{edition}{3rd} ed.

\bibitem[{\citenamefont{Brendel et~al.}(1991)\citenamefont{Brendel, Mohler, and
  Martienssen}}]{Brendel91}
\bibinfo{author}{\bibfnamefont{J.}~\bibnamefont{Brendel}},
  \bibinfo{author}{\bibfnamefont{E.}~\bibnamefont{Mohler}}, \bibnamefont{and}
  \bibinfo{author}{\bibfnamefont{W.}~\bibnamefont{Martienssen}},
  \bibinfo{journal}{Phys. Rev. Lett.} \textbf{\bibinfo{volume}{66}},
  \bibinfo{pages}{1142} (\bibinfo{year}{1991}).

\bibitem[{\citenamefont{{De Rossi} et~al.}(2001)\citenamefont{{De Rossi},
  Berger, Calligaro, Leo, Ortiz, and Marcadet}}]{DeRossi01}
\bibinfo{author}{\bibfnamefont{A.}~\bibnamefont{{De Rossi}}},
  \bibinfo{author}{\bibfnamefont{V.}~\bibnamefont{Berger}},
  \bibinfo{author}{\bibfnamefont{M.}~\bibnamefont{Calligaro}},
  \bibinfo{author}{\bibfnamefont{G.}~\bibnamefont{Leo}},
  \bibinfo{author}{\bibfnamefont{V.}~\bibnamefont{Ortiz}}, \bibnamefont{and}
  \bibinfo{author}{\bibfnamefont{X.}~\bibnamefont{Marcadet}},
  \bibinfo{journal}{Appl. Phys. Lett.} \textbf{\bibinfo{volume}{79}},
  \bibinfo{pages}{3758} (\bibinfo{year}{2001}).

\bibitem[{\citenamefont{Strekalov et~al.}(1996)\citenamefont{Strekalov,
  Pittman, Sergienko, Shih, and Kwiat}}]{Strekalov96}
\bibinfo{author}{\bibfnamefont{D.~V.} \bibnamefont{Strekalov}},
  \bibinfo{author}{\bibfnamefont{T.~B.} \bibnamefont{Pittman}},
  \bibinfo{author}{\bibfnamefont{A.~V.} \bibnamefont{Sergienko}},
  \bibinfo{author}{\bibfnamefont{Y.~H.} \bibnamefont{Shih}}, \bibnamefont{and}
  \bibinfo{author}{\bibfnamefont{P.~G.} \bibnamefont{Kwiat}},
  \bibinfo{journal}{Phys. Rev. A} \textbf{\bibinfo{volume}{54}},
  \bibinfo{pages}{R1} (\bibinfo{year}{1996}).

\end{thebibliography}
\end{document}